\documentstyle[manuscript,aps]{revtex}
\begin{document}


\title{Bound States of Non-Hermitian Quantum Field Theories}

\author{Carl M. Bender$^1$, Stefan Boettcher$^2$, H. F. Jones$^3$, Peter N.
Meisinger$^1$, and Mehmet \d{S}im\d{s}ek$^1$\cite{bye}}

\address{${}^1$Department of Physics, Washington University, St. Louis,
Missouri 63130, USA}

\address{${}^2$Department of Physics, Emory University, Atlanta, Georgia
30322, USA}

\address{${}^3$Blackett Laboratory, Imperial College, London SW7 2BZ, United
Kingdom}

\date{\today}

\maketitle

\begin{abstract}
The spectrum of the Hermitian Hamiltonian ${1\over2}p^2+{1\over2}m^2x^2+gx^4$
($g>0$), which describes the quantum anharmonic oscillator, is real and
positive. The non-Hermitian quantum-mechanical Hamiltonian $H={1\over2}p^2+{1
\over2}m^2x^2-gx^4$, where the coupling constant $g$ is real and positive, is
${\cal PT}$-symmetric. As a consequence, the spectrum of $H$ is known to be real
and positive as well. Here, it is shown that there is a significant difference
between these two theories: When $g$ is sufficiently small, the latter
Hamiltonian exhibits a two-particle bound state while the former does not. The
bound state persists in the corresponding non-Hermitian ${\cal PT}$-symmetric
$-g\phi^4$ quantum field theory for all dimensions $0\leq D<3$ but is not
present in the conventional Hermitian $g\phi^4$ field theory.
\end{abstract}


\vspace{.3in}
In this Letter we show that the spectrum of the non-Hermitian\footnote{Contrary
to appearances, this Hamiltonian is not Hermitian because its eigenfunctions are
required to obey boundary conditions in the complex plane [see (\ref{e2b})].}
${\cal PT}$-symmetric quartic Hamiltonian
\begin{eqnarray}
H={1\over2}p^2+{1\over2}m^2x^2-gx^4\qquad(g>0)
\label{e1}
\end{eqnarray}
possesses bound states for sufficiently small $g$. Moreover, when the Euclidean
space-time dimension $D$ satisfies $0\leq D<3$, the corresponding ${\cal
PT}$-symmetric quantum field theory described by the Lagrangian
\begin{equation}
{\cal L}={1\over2}(\nabla\phi)^2+{1\over2}m^2\phi^2-g\phi^4\qquad(g>0)
\label{e2}
\end{equation}
also possesses bound states for small $g$. The related conventional Hermitian
$gx^4$ and $g\phi^4$ theories with $g>0$ do not possess such bound states.

Quantum-mechanical Hamiltonians that are non-Hermitian but possess ${\cal PT}$
symmetry have recently been studied in great detail by many authors
\cite{X1,X2,X3,X4,X5,X6,X7,X8,X9,X10,X11,X12,X13,X14}. It is known that the
entire spectrum of the ${\cal PT}$-symmetric anharmonic oscillator is real and
positive. The first proof of the reality of the spectrum is given in
Ref.~\cite{X13}.
Direct numerical evidence for the reality and positivity of the spectrum can be
found by performing a Runge-Kutta integration of the associated Schr\"odinger
equation \cite{X1}. Alternatively, the large-energy eigenvalues of the spectrum
can be calculated with great accuracy by using conventional WKB techniques
\cite{X1,X3,BO}. An easy way to demonstrate the reality and positivity of the
spectrum is to calculate exactly the spectral zeta function (the sum of the
inverses of the eigenvalues). This was done by Mezincescu and Bender and Wang
\cite{X7,X9}.

The calculations that are performed in this paper are perturbative; that is,
quantities are determined as formal power series in the coupling constant $g$.
Some explanation must be given to justify the application of perturbative
methods to ${\cal PT}$-symmetric theories. We will argue that to obtain a
perturbation expansion for the theories described in (\ref{e1}) and (\ref{e2})
we need only calculate the perturbation series for the corresponding quantity in
the conventional Hermitian $gx^4$ and $g\phi^4$ theories and change the sign of
$g$ in these expansions. This procedure is correct but it is nontrivial, as we
now argue.

First, consider the quantum-mechanical theory in (\ref{e1}). The corresponding
differential equation for the energy eigenfunctions $\psi(x)$ is
\begin{eqnarray}
-{1\over2}\psi''(x)+{1\over2}m^2x^2\psi(x)-gx^4\psi(x)=E\psi(x)\qquad(g>0),
\label{e2a}
\end{eqnarray}
where the wave function $\psi(x)$ satisfies boundary conditions in two wedges of
$60^\circ$ opening angle in the complex-$x$ plane \cite{X1}:
\begin{eqnarray}
\lim_{|x|\to\infty}\psi(x)=0\qquad\left(-\pi/3<{\rm arg}\,x<0\quad{\rm and}\quad
-\pi<{\rm arg}\,x<-2\pi/3\right).
\label{e2b}
\end{eqnarray}
These two wedges lie just below the real axis. Because they are symmetric with
respect to the imaginary axis they enforce the ${\cal PT}$ symmetry of the
problem. In the limit as $g\to0$, the differential equation in (\ref{e2a})
becomes that of the harmonic oscillator and the boundary conditions are imposed
in two wedges of $90^\circ$ opening angle in the complex-$x$ plane:
\begin{eqnarray}
\lim_{|x|\to\infty}\psi(x)=0\qquad\left(-\pi/4<{\rm arg}\,x<\pi/4\quad{\rm and}
\quad-5\pi/4<{\rm arg}\,x<-{3\over4}\pi\right).
\label{e2c}
\end{eqnarray}
Because the wedges for $g>0$ and the wedges for $g=0$ have a region in common,
conventional perturbative methods are justified.

However, one must be extremely careful to avoid drawing wrong conclusions on the
basis of perturbative calculations. For example, a perturbative calculation
of the one-point Green's function (the expectation value of $x$ in the ground
state) gives $0$ to all orders in powers of the coupling constant $g$.
Nevertheless, the correct result is nonzero, pure imaginary, and exponentially
small \cite{BMY}:
\begin{eqnarray}
G_1\equiv{\int dx\,x\psi_0^2(x)\over\int dx\,\psi_0^2(x)}\sim
-i K m^{3/2}g^{-2/3}e^{-4m^3\over3g}\qquad(g\to0^+),
\label{e2d}
\end{eqnarray}
where $K=8\sqrt{\pi e}3^{-1/6}/\Gamma^2(1/3)$ is a positive constant.
The nonperturbative result in (\ref{e2d}) is due to a soliton solution of the
classical field equations.

There are even more subtle effects in which one may think that there are
nonperturbative effects when there are none. For example, consider the cubic
anharmonic oscillator Hamiltonian
\begin{eqnarray}
H={1\over2}p^2+{1\over2}m^2x^2+gx^3\qquad(g\,{\rm real}).
\label{e2e}
\end{eqnarray}
A perturbative calculation of the ground-state energy has the form
\begin{eqnarray}
E_0\sim{1\over2}m+\sum_1^\infty a_n(g^2)^n\qquad(g\to0).
\label{e2f}
\end{eqnarray}
The coefficients $a_n$ in this series are real and all have the {\it same sign}
and thus the series is not Borel summable. The nonsummability of this series
indicates the presence of a branch cut in the complex-$g$ plane on the real-$g$
axis. Thus, the function $E_0(g)$ is complex when $g$ is real. The imaginary
part of $E_0(g)$ is a nonperturbative effect; it is exponentially small when $g$
is real and small. The energies of the cubic oscillator are complex because of
tunneling effects on the real axis.

The corresponding ${\cal PT}$-symmetric cubic oscillator, whose Hamiltonian is
\begin{eqnarray}
H={1\over2}p^2+{1\over2}m^2x^2+igx^3\qquad(g\,{\rm real}),
\label{e2g}
\end{eqnarray}
is obtained from the Hamiltonian in (\ref{e2e}) by replacing $g$ with $ig$. This
replacement makes the series in (\ref{e2f}) Borel summable \cite{X5,X11} and
verifies the result that eigenvalues of $H$ in (\ref{e2g}) are all {\it real}.

However, this argument is quite tricky. Let us apply the same reasoning to the
conventional Hermitian quantum-mechanical anharmonic oscillator described by
\begin{eqnarray}
H={1\over2}p^2+{1\over2}m^2 x^2+gx^4\qquad(g>0).
\label{e3}
\end{eqnarray}
The Rayleigh-Schr\"odinger perturbation series for the ground-state energy of
this Hamiltonian has the form
\begin{eqnarray}
E_0\sim{1\over2}m+\sum_1^\infty a_ng^n\qquad(g\to0^+),
\label{e3a}
\end{eqnarray}
where the coefficients $a_n$ are all real and alternate in sign. Hence, this
series is Borel summable and we conclude (correctly) that the ground-state
energy is real when $g>0$. We correctly obtain the perturbation expansion for
the ground-state energy of the non-Hermitian ${\cal PT}$-symmetric anharmonic
oscillator (\ref{e1}) by replacing $g$ with $-g$. Now, the coefficients in
perturbation series for the ground-state energy no longer alternate in sign; the
coefficients all have the same sign (they are all negative). However, all of the
eigenvalues of the Hamiltonian (\ref{e1}) are real. Even though the series is
not Borel summable, the imaginary part of the ground-state energy is {\it
exactly zero} due to the presence of the soliton that leads to the
nonperturbative result in (\ref{e2d}).

The same discussion applies to the non-Hermitian ${\cal PT}$-symmetric quantum
field theory in (\ref{e2}). The perturbation expansions of the Green's functions
of this theory are obtained using the same Feynman rules as for the conventional
$g\phi^4$ quantum field theory, with the one change that the vertex amplitude
$-24g$ is replaced with $24g$. However, again one must be very careful about
nonperturbative effects. For example, the one-point Green's function, which
vanishes in the conventional $g\phi^4$ quantum field theory, is now
nonvanishing, pure imaginary, and exponentially small in the $-g\phi^4$ quantum
field theory \cite{BMY}.

Having made these preliminary remarks, we turn to the question of bound states
in ${\cal PT}$-symmetric theories. Let us first examine a two-particle state for
the conventional quantum anharmonic oscillator Hamiltonian (\ref{e3}).
The small-$g$ Rayleigh-Schr\"odinger perturbation series for the $k$th energy
level $E_k$ of this Hamiltonian is \cite{BWII}
\begin{eqnarray}
E_k\sim m\left[k+{1\over2}+{3\over4}(2k^2+2k+1)\epsilon+{\rm O}(\epsilon^2)
\right]\qquad(\epsilon\to0^+),
\label{e4}
\end{eqnarray}
where the dimensionless expansion parameter is $\epsilon=g/m^3$. In this theory
the {\it renormalized mass} $M$ is defined as the first excitation above the
ground state. Thus, from (\ref{e4}),
\begin{eqnarray}
M\equiv E_1-E_0\sim m\left[1+3\epsilon+{\rm O}(\epsilon^2)\right]\qquad
(\epsilon\to0^+).
\label{e5}
\end{eqnarray}

To determine if the two-particle state is bound, we examine the second
excitation above the ground state using (\ref{e4}). We define
\begin{eqnarray}
B_2\equiv E_2-E_0\sim m\left[2+9\epsilon+{\rm O}(\epsilon^2)\right]\qquad
(\epsilon\to0^+).
\label{e6}
\end{eqnarray}
If $B_2<2M$, then a two-particle bound state exists and the binding energy,
which is a negative quantity, is $B_2-2M$. If $B_2>2M$, then the second
excitation above the vacuum is interpreted as an unbound two-particle state. We
can see from (\ref{e6}) that in the small-coupling region, where perturbation
theory is valid, the conventional anharmonic oscillator does not possess a bound
state. Indeed, using a variety of methods (WKB theory, variational methods,
numerical calculations) one can show that there is no two-particle bound state
for any value of $g>0$. Because there is no bound state the $gx^4$ interaction
may be considered to represent a repulsive force.\footnote{In general, a
repulsive force in a quantum field theory is represented by an energy dependence
in which the energy of a two-particle state decreases with separation. The
conventional anharmonic oscillator Hamiltonian corresponds to a field theory in
one space-time dimension, where there cannot be any spatial dependence. In this
case the repulsive nature of the force is understood to mean that the energy
$B_2$ needed to create two particles at a given time is more than twice the
energy $M$ needed to create one particle.}

\begin{figure}
\vspace{5.0in}
\includegraphics{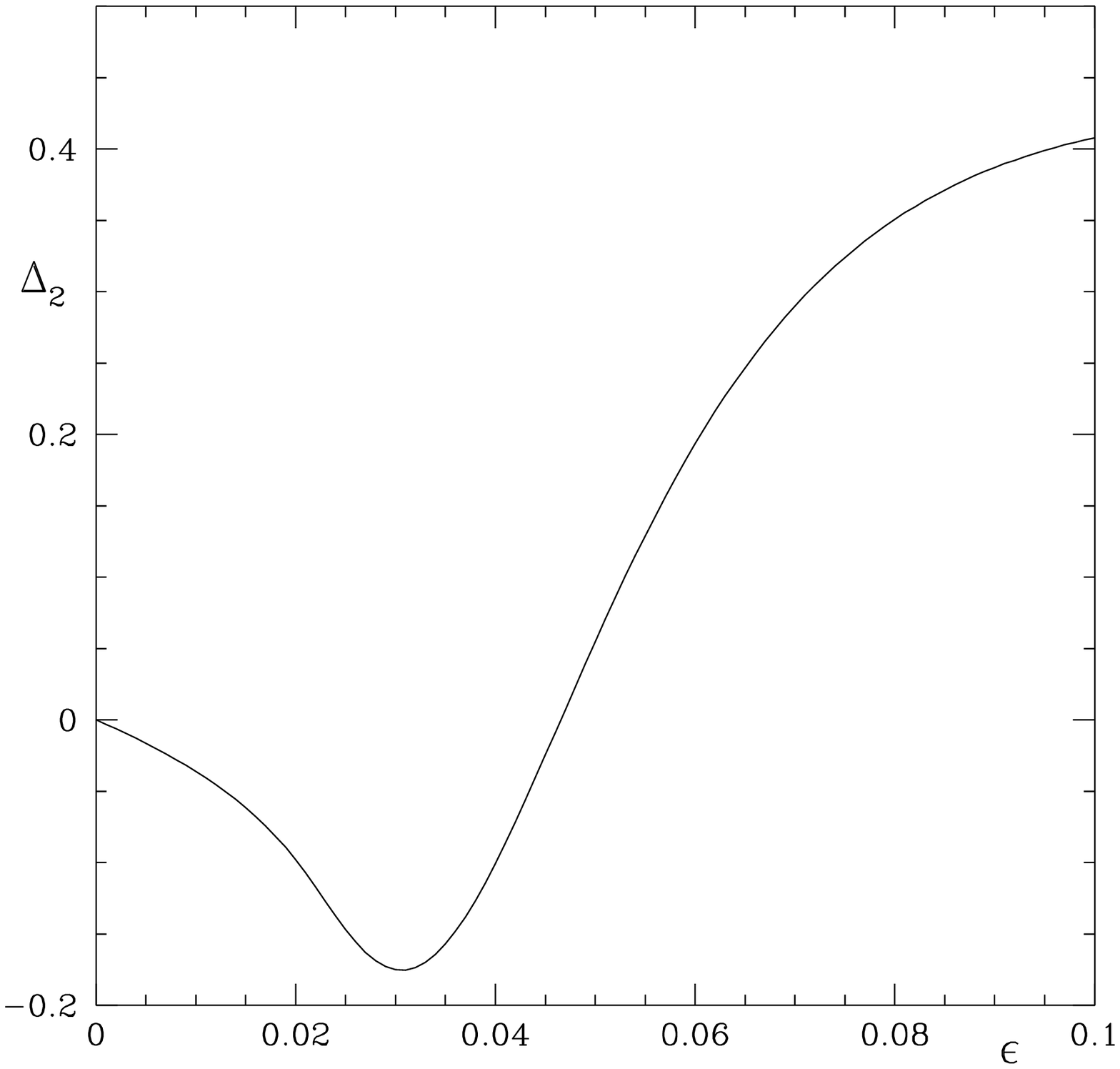}
\caption{Dimensionless binding energy $\Delta_2$ of the two-particle bound state
of the ${\cal PT}$-symmetric anharmonic oscillator, whose Hamiltonian $H$ is
given in Eq.~(1), plotted as a function of the dimensionless coupling constant
$\epsilon=g/m^3$. The quantity $\Delta_2$ represents the binding energy $B_2$
measured in units of the renormalized mass [see Eq.~(15)]. Note that for small
$\epsilon$ the slope of the curve is $-3$, which verifies the asymptotic result
in (15). Observe that the bound state disappears when $\epsilon$ increases
beyond $0.0465\ldots$. As $\epsilon$ continues to increase, $\Delta_2$ reaches
a maximum value of $0.427$ at $\epsilon=0.13$. Then $\Delta_2$ decreases and
approaches the limiting value $0.28$ as $\epsilon\to\infty$.}
\label{f1}
\end{figure}

We obtain the perturbation series for the ${\cal PT}$-symmetric oscillator,
whose Hamiltonian $H$ is given in (\ref{e1}), from the perturbation series for
the conventional anharmonic oscillator by replacing $\epsilon$ with $-\epsilon$.
Thus, while the conventional anharmonic oscillator does not possess a
two-particle bound state, evidently the ${\cal PT}$-symmetric oscillator does
indeed possess such a state. The binding energy of this state can be measured in
units of the renormalized mass $M$. For this purpose we define the {\it
dimensionless} binding energy $\Delta_2$ by
\begin{eqnarray}
\Delta_2\equiv{B_2-2M\over M}\sim-3\epsilon+{\rm O}(\epsilon^2)\qquad(\epsilon
\to0^+).
\label{e7}
\end{eqnarray}
In Fig.~\ref{f1}, $\Delta_2$ is plotted as a function of the dimensionless
coupling constant $\epsilon$. Observe that the bound state disappears when the
value of $\epsilon$ increases beyond $\epsilon=0.0465\ldots$. As $\epsilon$
continues to increase, $\Delta_2$ reaches a maximum value of $0.427$ at
$\epsilon=0.13$ and then approaches the limiting value $0.28$ as $\epsilon\to
\infty$.

It is interesting that the bound state for the ${\cal PT}$-symmetric anharmonic
oscillator occurs only for values of $\epsilon$ less than about $0.0465$. The
following heuristic argument helps to explain why the characteristic size of
$\epsilon$ is so small. For the Hamiltonian in (\ref{e1}) the potential has the
form $V(x)={1\over2}m^2x^2-gx^4$. On the real-$x$ axis, the maximum value of
$V(x)$ occurs at $x^2={1\over4}m^2/g$ and at this point $V_{\rm max}={1\over16}
m^4/g$. In order to have a two-particle bound state it is necessary for the
height of the potential to be high enough to bind three states, the largest of
which is of order ${5\over2}m$. Setting this value equal to $V_{\rm max}$ gives
$\epsilon={1\over40}$, which is near the value for which the two-particle state
is most strongly bound. Note that the potential $V(x)$ does not allow states to
tunnel out of the well, even though the potential decreases for large real $x$.
This is because the ${\cal PT}$-symmetric boundary conditions are not imposed on
the real axis but rather in a pair of wedges in the complex-$x$ plane that lie
below the real axis [see Eq.~(\ref{e2b})].

\begin{figure}
\vspace{5.0in}
\includegraphics{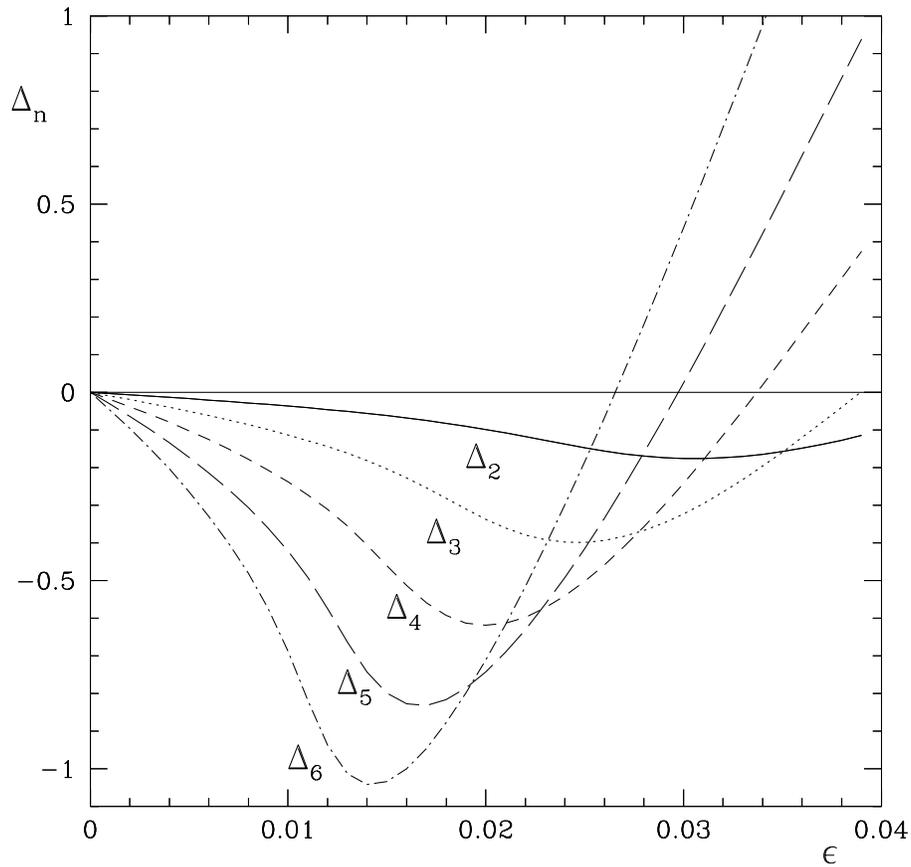}
\caption{Dimensionless binding energies $\Delta_2$, $\Delta_3$, $\Delta_4$,
$\Delta_5$, and $\Delta_6$ for the two-, three-, four-, five-, and six-particle
bound states plotted as functions of the dimensionless coupling constant
$\epsilon$. Note that the multiparticle bound states cease to be bound as
$\epsilon$ increases past $0.039$, $0.034$, $0.030$, and $0.027$.}
\label{f2}
\end{figure}

In the ${\cal PT}$-symmetric anharmonic oscillator, there are not only
two-particle bound states for small coupling constant but also $k$-particle
bound states for all $k\geq2$. To see why, we calculate the energy of the $k$th
excitation above the vacuum:
\begin{eqnarray}
B_k\equiv E_k-E_0\sim m\left[k-{3\over2}(k^2+k)\epsilon+{\rm O}(\epsilon^2)
\right]\qquad(\epsilon\to0+),
\label{e8}
\end{eqnarray}
which is the generalization of (\ref{e6}). Thus, the dimensionless binding
energy is
\begin{eqnarray}
\Delta_k\equiv{B_k-kM\over M}\sim-{3\over2}k(k-1)\epsilon+{\rm O}(\epsilon^2)
\qquad(\epsilon\to0+),
\label{e9}
\end{eqnarray}
This equation reduces to (\ref{e7}) when $k=2$. The key feature of this equation
is that the coefficient of $\epsilon$ is negative. Since the dimensionless
binding energy becomes negative as $\epsilon$ increases from $0$, there is a
$k$-particle bound state. The dimensionless binding energies for the first
five $k$-particle bound states are shown in Fig.~\ref{f2}. Note that the higher
multiparticle bound states cease to be bound for smaller values of $\epsilon$;
starting with the three-particle bound state, the binding energy of these states
becomes positive as $\epsilon$ increases past $0.039$, $0.034$, $0.030$, and
$0.027$. Furthermore, the heuristic argument given above for the value of
$\epsilon$ at which the multiparticle states are most strongly bound is
qualitatively correct; one can see that the $k$-particle bound state is most
strongly bound at a value of $\epsilon$ that is roughly $1/(16k+8)$.

Figure \ref{f2} shows that for any value of $\epsilon=g/m^3$ there are always a
finite number of bound states and an infinite number of unbound states. The
number of bound states decreases with increasing $\epsilon$ until, as we see in
Fig.~\ref{f1}, there are no bound states at all. Observe that there is a range
of $\epsilon$ for which there are only two- and three-particle bound states.
(This situation is analogous to the physical world in which one observes only
states of two and three bound quarks.) In this range of $\epsilon$ if one has an
initial state containing a number of particles (renormalized masses), these
particles will clump together into bound states, releasing energy in the
process. Depending on the value of $\epsilon$, the final state will consist
either of two- or of three-particle bound states, whichever is energetically
favored. Note also that there is a special value of $\epsilon$ for which two-
and three-particle bound states can exist in thermodynamic equilibrium.

Let us now generalize the result for the quantum-mechanical two-particle bound
state in (\ref{e7}) to the case of the $D$-dimensional ${\cal PT}$-symmetric
quantum field theory in (\ref{e2}). To show that there is a bound state, we need
only demonstrate that {\it to leading order in the coupling constant}, the
binding energy becomes negative as the coupling constant increases from $0$. The
Feynman rules for the calculation are the conventional ones, except that the
sign of the vertex is reversed: $24g$ for a four-point vertex amplitude and $1/
(p^2+M^2)$ for a line amplitude. Note that to leading order in the coupling
constant we use the renormalized mass $M$ rather than the unrenormalized mass
$m$ in the propagator amplitude and we exclude self-energy loops in the graphs.

\begin{figure}
\vspace{5.0in}
\includegraphics{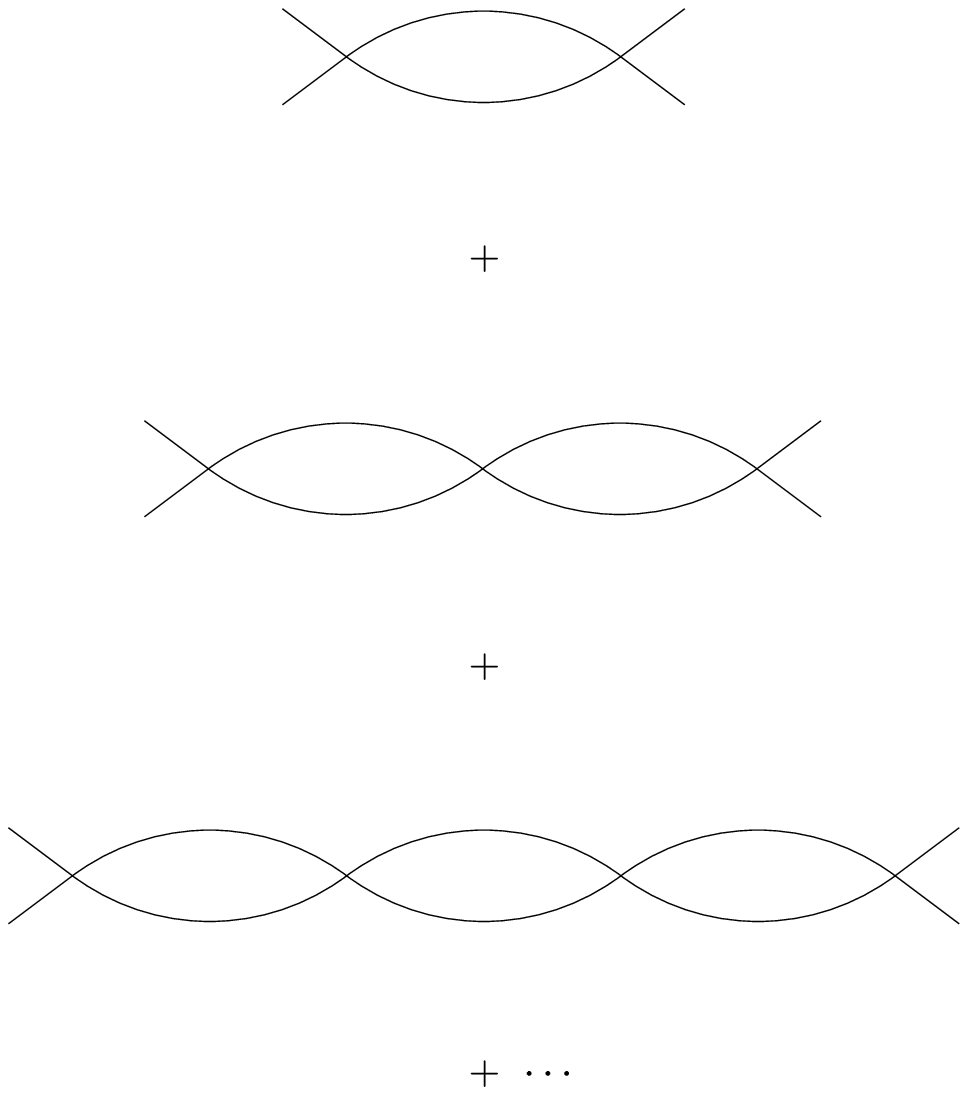}
\caption{So called ``sausage-link'' graphs contributing to the leading-order
value of the energy of the two-particle bound state for the ${\cal
PT}$-symmetric $-g\phi^4$ quantum field theory whose Euclidean Lagrangian is
given in Eq.~(2).}
\label{f3}
\end{figure}

To calculate the energy of the bound state to leading order we sum up all
``sausage-link'' graphs and identify the bound-state pole in the geometric
series. These graphs are shown in Fig.~\ref{f3}. The result for the
dimensionless binding energy to leading order in the dimensionless coupling
constant
$\epsilon=gM^{D-4}$ is \begin{eqnarray}
\Delta_2\sim-4\left[
{3\Gamma\left({3-D\over2}\right)\over4\pi^{(D-1)/2}}\right]^{2/(3-D)}
\epsilon^{2/(3-D)},
\label{e10}
\end{eqnarray}
where $D$ is the Euclidean space dimension. The formula reduces to that in
(\ref{e7}) for the case $D=1$. This formula is valid for $0\leq D<3$. Note that
we have performed a mass renormalization to leading order, but we have not done
a wave function or coupling-constant renormalization in this calculation.

We conclude by remarking that the behavior of a $g\phi^3$ theory is the reverse
of that of a $g\phi^4$ theory. A $g\phi^3$ theory represents an attractive
force. The bound states that arise as a consequence of this force can be found
by using the Bethe-Salpeter equation. However, the $g\phi^3$ field theory is
unacceptable because the spectrum is not bounded below. If we replace $g$ by
$ig$, the spectrum now becomes real and positive, but the force becomes
repulsive and there are no bound states. The same is true for a two-scalar
theory with interaction of the form $ig\phi^2\chi$. This latter theory is an
acceptable model of scalar electrodynamics, but has no analog of positronium.

\section*{ACKNOWLEDGMENTS}
\label{s1}
M\d{S} is grateful to the Physics Department at Washington University for their
hospitality during his sabbatical. This work was supported by the
U.S.~Department of Energy.

\newpage
\center{FIGURE 2}

\newpage
\center{FIGURE 3}

\begin{references}

\bibitem[*]{bye} Permanent address: Gazi Universitesi, Fen Edebiyat Fakultesi,
Fizik Bolumu, 06500 Teknikokullar-Ankara, Turkey.

\bibitem{X1} C.~M.~Bender and S.~Boettcher,
Phys.~Rev.~Lett.~{\bf 80}, 5243 (1998).  

\bibitem{X2} E.~Delabaere and F.~Pham,
Phys.~Lett.~A {\bf 250}, 25 (1998)   
and   
29 (1998).  

\bibitem{X3} C.~M.~Bender, S.~Boettcher, and P.~N.~Meisinger,
J.~Math.~Phys. {\bf 40}, 2201 (1999).  

\bibitem{X4} C.~M.~Bender, F.~Cooper, P.~N.~Meisinger, and V.~M.~Savage,
Phys.~Lett.~A {\bf 259}, 224 (1999).

\bibitem{X5} C.~M.~Bender and G.~V.~Dunne,
J.~Math.~Phys.\ {\bf 40}, 4616 (1999).

\bibitem{X6} E.~Delabaere and D.~T.~Trinh,
J.~Phys.~A: Math.~Gen. {\bf 33}, 8771 (2000).  

\bibitem{X7} G. A. Mezincescu,
J.~Phys.~A: Math.~Gen.~{\bf 33}, 4911 (2000).  

\bibitem{X8} C. M. Bender, S. Boettcher, and V. M. Savage,
J.~Math.~Phys. {\bf 41}, 6381 (2000).  

\bibitem{X9} C.~M.~Bender and Q.~Wang,
to be published in J.~Phys.~A: Math.~Gen.

\bibitem{X10} K.~C.~Shin,
University of Illinois preprint.

\bibitem{X11} C.~M.~Bender and E.~J.~Weniger, J.~Math.~Phys.~{\bf 42}, 2167-2183
(2001).

\bibitem{X12} C. M. Bender, G. V. Dunne, P. N. Meisinger, and M. \d{S}im\d{s}ek,
Phys.~Lett.~A~{\bf 281}, 311-316 (2001).

\bibitem{X13} P.~Dorey, C.~Dunning, and R.~Tateo, hep-th/0103051.

\bibitem{X14} C.~M.~Bender, M.~Berry, P.~N.~Meisinger, V.~M.~Savage, and
M.~\d{S}im\d{s}ek, J.~Phys.~A: Math.~Gen.~{\bf 34}, L31-L36 (2001).

\bibitem{BO} C.~M.~Bender and S.~A.~Orszag, {\it Advanced Mathematical Methods
for Scientists and Engineers} (McGraw-Hill, New York, 1978), Chap.~10.

\bibitem{BMY} C.~M.~Bender, P.~Meisinger, and H.~Yang, Phys.~Rev.~D {\bf 63},
45001 (2001).

\bibitem{BWII} C.~M.~Bender and T.~T.~Wu,
Phys.~Rev.~D {\bf 7}, 1620 (1973), Eq.~(5.11).

\end{references}
\end{document}